# Quelle éthique pour quelle IA ?[1]

*David Doat*

Maître de conférences en Philosophie
Equipe « Ethique, Technologie & Humanités » (ETH+)
UR-ETHICS EA 7446
Université Catholique de Lille

Mots-clés : *éthique, intelligence artificielle, théories morales*

Résumé : Cette étude propose une analyse des formes d'éthiques impliquées en éthique de l'IA, et en situe les intérêts et les limites. Dans un premier temps, l'auteur introduit au besoin contemporain et au sens de l'éthique. Il la distingue d'autres registres de normativités et en souligne le caractère impropre à la formalisation. Il présente ensuite une cartographie du paysage des théories éthiques que recouvre la philosophie morale, en prenant soin de distinguer la méta-éthique, l'éthique normative et l'éthique appliquée. En dressant ce panorama, l'auteur interroge les relations entre l'éthique et l'intelligence artificielle. L'analyse porte en particulier sur les grands courants de l'éthique qui se sont imposés dans les façons de faire de l'éthique du numérique et de l'IA dans nos démocraties occidentales. L'auteur se demande si ces pratiques, telles qu'elles semblent aujourd'hui cristallisées en une configuration précise, constituent une réponse suffisante et suffisamment satisfaisante à nos besoins d'éthique en matière d'IA. L'étude s'achève par une réflexion sur les raisons pour lesquelles une éthique humaine de l'IA, fondée dans une pratique pragmatiste de l'éthique en situation, reste nécessaire et irréductible à toute entreprise de formalisation ou de traitement automatisé des questions éthiques qui se posent aux humains.

Summary: This study proposes an analysis of the different types of ethical approaches involved in the ethics of AI, and situates their interests and limits. First, the author introduces to the contemporary need for and meaning of ethics. He distinguishes it from other registers of normativities and underlines its inadequacy to formalization. He then presents a cartography of the landscape of ethical theories covered by moral philosophy, taking care to distinguish meta-ethics, normative ethics and applied ethics. In drawing up this overview, the author questions the relationship between ethics and artificial intelligence. The analysis focuses in particular on the main ethical currents that have imposed themselves in the ways of doing digital ethics and AI in our Western democracies. The author asks whether these practices of ethics, as they seem to crystallize today in a precise pattern, constitute a sufficient and sufficiently satisfactory response to our needs for ethics in AI. The study concludes with a reflection on the reasons why a human ethics of AI based on a pragmatic practice of contextual ethics remains necessary and irreducible to any formalization or automated treatment of the ethical questions that arise for humans.

---

[1] Les bases de la présente étude furent posées par l'auteur à l'occasion d'une intervention au sein du Workshop « Ethique et morale » de la Chaire IA Responsable de l'Université d'Artois (France), le 28 mai 2021.

# Plan

## Introduction

Nous proposons dans la présente étude une analyse des formes d'éthiques impliquées en éthique de l'intelligence artificielle (IA), et en situons les intérêts et les limites. Dans un premier temps, nous introduisons au besoin contemporain et au sens de l'éthique que nous distinguons d'autres registres de normativités, et soulignons son caractère impropre à la formalisation. Nous présentons ensuite une cartographie du paysage des théories éthiques que recouvre la philosophie morale, en prenant soin de distinguer la méta-éthique, l'éthique normative et l'éthique appliquée. En dressant ce panorama, nous interrogeons les relations entre l'éthique et l'intelligence artificielle. Notre analyse porte en particulier sur les grands courants de l'éthique qui se sont imposés dans les façons de faire de l'éthique du numérique et de l'IA dans nos démocraties occidentales. Nous nous demandons si ces pratiques, telles qu'elles semblent aujourd'hui cristallisées en une configuration précise, constituent une réponse suffisante et suffisamment satisfaisante à nos besoins d'éthique en matière d'IA. Nous concluons notre étude par une réflexion sur les raisons pour lesquelles une éthique humaine de l'IA, fondée dans une pratique pragmatiste de l'éthique en situation, reste nécessaire et irréductible à toute entreprise de formalisation ou de traitement automatisé des questions éthiques qui se posent aux humains.

# 1. Les facteurs d'un besoin d'éthique contemporain

Nous vivons une époque où s'exprime un grand besoin d'éthique à tous les niveaux de la société, et le champ de l'intelligence artificielle (IA) n'y échappe pas. Cette situation de besoin est le résultat de nombreux facteurs. Nous pouvons en souligner trois en particulier, sans aucune visée d'exhaustivité.

Un premier facteur d'intérêt et de besoin d'éthique aujourd'hui est lié au rythme soutenu de l'évolution technologique. Nous évoluons dans des sociétés complexes, de plus en plus conditionnées par des technologies qui nous donnent de nouveaux pouvoirs, mais nous posent en même temps d'innombrables questions nouvelles sur nos responsabilités, nos droits et nos devoirs, questions auxquelles les religions et les sagesses morales du passé n'apportent pas de réponse évidente suffisamment précise par rapport aux situations concrètes où nous plongent nos nouvelles technologies et nos nouveaux pouvoirs. Nous sommes donc aculés à devoir innover en matière d'éthique dans tous les domaines où l'évolution technologique vient déstabiliser nos systèmes éthiques traditionnels[2].

Un second facteur est celui de la mise en question de la thèse de l'exception humaine : on a toujours vu en l'homme un être à part, doué de capacités rationnelles qui en faisaient un être d'exception sans égal dans la nature. Or, depuis à peu près deux siècles, la thèse de l'exception humaine[3] n'a cessé d'être érodée et battue en brèche (darwinisme, éthologie et primatologie, informatique, IA…). Entre IA, robots et animaux, y a-t-il encore un propre de l'homme ? Dans un monde où la spécificité humaine est mise en question de part et d'autre de ses deux bords (animalité, machine), la dimension éthique de l'existence humaine peut encore apparaître comme un phare dans la nuit, un lieu fort où s'exprime la différence anthropologique.

Un troisième facteur de nos besoins contemporains d'éthique est lié au caractère pluraliste de nos sociétés contemporaines, et aux effets de la mondialisation dans nos pays démocratiques en particulier. Il n'est plus possible aujourd'hui de se référer universellement à une religion ou une tradition morale unique pour juger de ce qui est bien ou mal, juste ou injuste. Les sources morales, philosophiques et religieuses de nos sociétés démocratiques se sont très largement pluralisées en Occident au cours d'un long processus historique, qu'a très bien analysé un philosophe comme Charles Taylor dans *Les sources du moi*[4]. Dans un contexte pluraliste, sécularisé et pluri-convictionnel, la référence à une tradition morale unique pour guider l'agir humain n'est plus possible. Notons que l'on pourrait sans doute nuancer ce propos, tant il apparaît que la Déclaration universelle des droits de l'homme de 1946 constitue pour les sociétés démocratiques et les Etats de droit, un cadre de référence juridique et éthique partagé, tout particulièrement dans le cadre des politiques de l'Union Européenne. Ceci étant dit, notre contexte occidental reste marqué par son caractère pluraliste, sécularisé et pluriconvictionnel. L'éthique y est favorablement accueillie comme une démarche de recherche du juste et du bien qui paraît d'emblée plus appropriée à notre situation sociale et politique contemporaine, si l'on distingue l'éthique de la morale par le fait que la première est davantage un processus réflexif dénué de certitudes *a priori*, qu'un système de convictions et de certitudes déjà établi (c'est-à-dire un système moral), et souvent rattaché à certaines visions philosophiques ou religieuses du monde et des rapports humains.

---

[2] J. Ladrière, *l'éthique dans l'univers de la rationalité*, Montréal, Fides ; Namur, Artel, 1997.
[3] J.-M. Schaeffer, *La fin de l'exception humaine*, Paris, Gallimard 2007.
[4] C. Taylor, *Les sources du moi*, Paris, Seuil, 1998.

## 2. Contenu et sens premier de l'éthique

Tout ceci étant dit, de quelle éthique a-t-on besoin aujourd'hui pour faire de l'IA? Selon Thierry Ménissier, philosophe, *« [à] l'heure actuelle, en dépit des apparences, personne ne peut dire avec certitude quelle forme d'éthique est exactement adéquate au traitement approfondi des situations engendrées par le développement de l'IA dans la société. Clore trop rapidement ce débat pourrait même conduire à un appauvrissement préjudiciable à la cause que l'on veut promouvoir grâce à lui, à savoir, redonner voix, dans un monde en cours d'automatisation exponentielle, à un humain à la fois actif et responsable[5]. »* Mais qu'est-ce que l'éthique ? Son sens peut être recherché en la distinguant de la morale, mais aussi en cherchant à dégager les bases naturelles d'un « sens moral », héritées de notre condition biologique, éthologique et évolutionnaire, telles que l'empathie, la réciprocité, ou le souci de la communauté et de la bonne entente en son sein, ainsi que ses conditions psychologiques et cognitives. Nous pouvons aussi souligner le rôle crucial des normes dans la socialité humaine, l'intérêt d'instituer une pratique de l'éthique dans les organisations de santé ou les entreprises, et la pluralité des sources culturelles de l'éthique qui invite à ne pas s'enclore dans des systèmes de pensée strictement occidentaux en matière de réflexion éthique sur l'IA.

Toujours selon Thierry Ménissier, *« [la] démarche éthique consiste à délimiter la responsabilité humaine et à permettre aux humains de savoir comment orienter leur liberté. Aussi, revendiquer l'éthique revient, pour un sujet, à assumer l'ambition d'être normatif. Est normatif ce qui émet des jugements de valeur, institue des règles et des principes d'action. L'ambition éthique est spécifique, car elle s'inscrit dans la perspective de la recherche d'une normativité qui n'est ni celle de la loi, ni celle de la coutume, ni celle de la déontologie d'une profession ou d'une corporation : si les diverses normativités, qu'elles soient juridique, sociologique ou éthique, peuvent parfois concorder, elles peuvent également diverger voire s'opposer[6]. »*. Ce non-alignement des normativités peut être source d'ambiguïtés, d'incertitudes voire de contradictions qui confèrent à la démarche éthique, de façon tout à fait explicite dans ces cas de figure, une fonction de jugement ultime et de quête de solutions nouvelles, dans des situations où ces solutions ne peuvent être produites sur la base de simples déductions ou de procédures de résolution établies.

De ce point de vue, l'éthique ne se définit pas par un champ de normes qui lui seraient spécifiques, radicalement hétérogènes aux registres des normes sociales, juridiques ou déontologiques. Agir de façon éthique, mener une démarche éthique, conduire une analyse éthique, ou encore émettre une évaluation éthique ne se réduit pas à appliquer mécaniquement sur un objet un référentiel de principes étrangers à la morale, aux mœurs ou au droit, dont on présupposerait qu'ils seraient la matière même de l'éthique, projetés sur cet objet pour en mesurer l'acceptabilité éthique. Agir de façon éthique, c'est exercer avec, ou *a minima* sous le contrôle de toutes les parties prenantes, et au moyen de tout instrument éventuel, une capacité d'appréciation, d'interprétation et d'application, mais aussi parfois de subversion ou de création de normes adaptées aux enjeux des situations qui appellent des décisions ou des régulations à même d'orienter l'action selon ce qui apparaîtra, à toutes et tous comme le plus juste et le meilleur – dans des situations, il faut insister, où ces résolutions ne peuvent être produites sur la base de simples déductions ou de procédures de résolution évidentes.

---

[5] T. Ménissier, « Quelle éthique pour l'IA ? », Naissance et développements de l'intelligence artificielle à Grenoble, Académie Delphinale, octobre 2019, Grenoble, France, lien internet : https://halshs.archives-ouvertes.fr/halshs-02398215/document. Ma présente réflexion doit une partie de son inspiration à la lecture de cette conférence stimulante donnée par Thierry Ménissier en 2019 à Grenoble.
[6] *Idem*.

Pour préciser notre propos, nous soulignons deux dimensions de l'éthique en recourant à une analogie. De même que l'on différencie la succession des images qui composent la continuité d'un film, du film lui-même dont le mouvement ne s'arrête jamais sur l'une de ses images particulières, on peut aussi distinguer les systèmes de règles et de principes (sociaux, juridiques, déontologiques ou moraux) qui, dans des cadres historiques et culturels divers, codifient les comportement humains, de la capacité humaine de produire de tels systèmes, de les juger et d'en changer, quelles que soient leurs différences historiques et culturelles. Nous disposons en tant qu'être humain d'un pouvoir dynamique de jugement et de création de normes, qu'elles soient morales, juridiques, déontologiques ou sociales. Mais ce pouvoir s'accompagne aussi d'une capacité d'interprétation, de réforme et de subversion normative qui ne doit pas être confondue avec tous systèmes particuliers de règles et de principes, lesquels ne sont jamais que les conséquences de ce pouvoir, les trâces laissées dans le temps et l'histoire par ce que Jean Ladrière, éminent philosophe belge du siècle dernier, appelle la « dimension éthique de l'existence[7] » humaine. Car de même que la grandeur d'un film ne dépend pas d'abord de ses images arrêtées, mais de l'inspiration et des qualités de son ou ses auteurs, de même, notre pouvoir individuel et collectif d'interprétation, de subversion ou de création normative nous responsabilise au plus haut point quant aux sens et aux finalités de nos actions, dès lors qu'il dépend de nous, de façon radicale, d'instaurer et de nous porter garants des conditions normatives d'une vie bonne, avec et pour autrui, dans des institutions justes - pour reprendre ici la définition de l'éthique de Paul Ricoeur. De ce point de vue, la morale, le droit et nos codes de déontologie sont à l'éthique, comme dimension fondamentale de l'existence humaine, ce que les images d'un film sont à sa dynamique, ou sa mobilité : un film ne peut se passer de ses images, mais il ne se réduit pas à ces dernières. Chacune porte au contraire la trace d'un sens et d'une histoire dont elles dépendent et qui les dépasse. De même, sans s'y épuiser, l'éthique ne peut se passer de la morale, du droit, de la déontologie, de nos mœurs et usages sociaux ; inversement, si ceux-ci ne se réduisent pas à l'éthique, cette dernière les éclaire en leur conférant un sens, en les inscrivant dans une histoire.

Reste néanmoins la question suivante : notre capacité d'évaluer, modifier, suivre ou non certaines normes quand nous le jugeons nécessaire, suit-elle elle-même certaines normes ? La question n'est pas anodine, notamment si l'on souhaite produire une IA éthique, capable de raisonnement éthique. Il faudrait en effet pouvoir objectiver et traduire sous forme d'algorithmes les règles, s'il en existe, qui procéduraliseraient notre capacité de juger, subvertir ou inventer de nouvelles règles. Juger ou agir de façon éthique supposent en effet parfois de sortir du cadre normatif usuel d'un milieu, d'une époque ou d'une culture donnée, pour proposer de façon transgressive de nouveaux critères d'évaluation de l'action, de nouvelles façons d'agir et de se comporter. Cette capacité de distanciation critique, parfois de transgression et d'invention de nouvelles règles dépend-elle elle-même de certaines règles, qui, une fois mises à jour, pourraient être algorithmisées ? Pourrait-on, pour le dire autrement, inventer un algorithme qui soit capable de réécrire ses procédures, de s'émanciper de son code profond, et de se donner de nouvelles règles de fonctionnement ? Cette question rejoint celle du problème de l'invention, et fait l'objet de nombreux débats, y compris dans des sciences apparemment aussi éloignées de l'éthique que les mathématiques. En effet, si l'invention mathématique est une chose bien connue des mathématiciens, l'on ne dispose pas encore d'une théorie mathématique de l'invention. De même, si l'évaluation et la recherche de solutions nouvelles face à des problèmes complexes (relevant par exemple de l'ordre du dilemme, du conflit de valeurs, etc.) sont des choses bien connues des éthiciens, ses conditions d'émergence ne permettent pas d'en dégager des invariants. Tout porte à croire au contraire que l'exercice du jugement et l'inventivité en éthique soient éminemment singuliers, dépendants des situations, non prédictibles, non systématisables et que leur mode opératoire résiste à toute tentative de formalisation.

---

[7] Voir J. Ladrière, *L'éthique dans l'univers de la rationalité*, *op. cit.*.

### 3. Petite cartographie du paysage des théories éthiques

Après avoir suggéré un certain sens et contenu à l'éthique, nous proposons de présenter dans leurs grandes lignes les théories éthiques que recouvre le paysage classique de la philosophie morale. Par facilité d'usage, nous considérerons dans ce contexte les termes « éthique » et « philosophie morale » dans leur sens très général, pour désigner un type de pensée visant à établir des relations entre un certain sens du bien ou du juste qu'un sujet, un collectif ou une société associe à ses décisions ou à ses actions, et les règles qui en découlent. Nous nous éloignons donc de l'acception plus spécifique de l'éthique que nous avons posée précédemment, pour y revenir dans notre conclusion.

Un mot, tout d'abord, sur l'étendue du propos : le champ institutionnel et académique de la philosophie morale s'est historiquement constitué, depuis l'Antiquité, dans les écoles grecques et les universités occidentales. Si bien des raisons justifieraient que l'on s'intéresse aussi aux éthiques asiatiques, orientales ou africaines, tel n'est toutefois pas l'objectif de cette étude. En effet, l'horizon de la tradition occidentale permet déjà d'apercevoir, comme nous le verrons, que l'éthique de l'IA, comme elle se présente de façon dominante aujourd'hui, est le résultat d'un certain nombre de réductions, de sélections plus ou moins conscientes et de mises entre parenthèses de pans entiers de la sagesse et de la richesse de la culture occidentale en matière d'éthique et de réflexion morale. Tant et si bien que faire de l'éthique de l'IA aujourd'hui, dans ses formes déterminées par la pensée universitaire, la pensée du droit et l'ordre des normes institutionnelles (soft law), c'est à la fois une démarche excellente et tout à fait nécessaire, mais c'est aussi, bien souvent, une entreprise réductrice, qui fait entrer dans un cadre qui peut conduire, parfois, à des apories[8].

Ce point précisé, de quelle éthique a-t-on besoin aujourd'hui pour faire de l'IA? La question est complexe, car il y a diverses manières de faire de l'éthique, comme en atteste la philosophie morale, où l'on peut distinguer le champ de la méta-éthique ou des fondements ultimes de l'éthique, le champ de l'éthique normative et celui de l'éthique appliquée ou éthique dite pratique.

### *3.1. La méta-éthique*

Le champ de la méta-éthique remonte à l'Antiquité. Chez Platon, l'idée de Justice, comme l'idée du Bien, sont des concepts en soi évidents pour tout esprit humain, bien que ces concepts n'aient pas de représentations sensibles - de même qu'il n'y a pas de traduction empirique d'un cercle parfait dans la nature, bien que nous en ayons une idée claire et distincte, personne n'a jamais eu d'expérience de la Justice en soi ; il en va de même du Bien. Les religions monothéistes l'identifient à Dieu, mais « Dieu, personne ne l'a jamais vu ». Ainsi que ce soit en philosophie ou en religion, la Justice comme le Bien sont en quelque sorte des idées *a priori*, des transcendantaux, comme disent les philosophes, ou deux horizons qui nous échappent toujours ; nous en avons un certain sens, des intuitions et des sentiments profonds (parfois dans l'expérience de leur contraire, le sentiment d'injustice ou de mépris en étant deux exemples très concrets parmi d'autres), et nous évaluons moralement les situations humaines à partir de leur étalon idéal, que nous trouvons en nous comme en tout homme doué de raison ; cette idée que les sens du bien et du juste font en quelque sorte partie d'une structure *a priori* de l'esprit humain, ou de ce à quoi l'esprit humain nous permet d'accéder, va courir tout au long de l'histoire de la philosophie, de Platon à Kant, jusqu'à aujourd'hui. Les approches empiristes, naturalistes ou évolutionnaires de l'éthique (comme celle de l'éthologue Frans de Waal[9]), vont critiquer et amender cette tradition de pensée, en soutenant que nos sens du bien et du juste sont plutôt des constructions historiques biologiquement et socio-culturellement déterminées.

Ces débats virulents entre idéalistes et naturalistes relèvent d'un champ de l'éthique en particulier : le champ de la méta-éthique. La méta-éthique est à l'éthique, ce que la méta-physique est à la physique, pourrait-on dire. La méta-éthique s'intéresse aux fondements de nos concepts moraux, à leurs origines et à leur justification, ainsi qu'au sens et aux justifications ultimes de nos actions. Pourquoi faisons-nous ce que nous faisons ? Quelles sont les finalités poursuivies, les conceptions dominantes du bien et du juste qui président à nos choix de société[10] ? Sont-elles cohérentes, sont-elles justifiées ? Etc. Par rapport à l'IA,

---
[8] Cette remarque ne se limite pas à l'éthique de l'IA, elle vaut aussi pleinement pour les champs de l'éthique de la donnée et de l'éthique du numérique.
[9] F. de Waal, *Le bon singe. Les bases naturelles de la morale,* Paris, Bayard, 1997.
[10] Notons que le fait qu'il y ait désaccord sur ces questions, de même que le fait que nous soyons capables de faillir, d'erreurs cognitives et morales, ou d'agir contrairement à nos intuitions morales, constitue paradoxalement autant de conditions nécessaires de la moralité ou de l'éthicité humaine. Si tout était clair et que nous agissions toujours nécessairement selon le juste et le bien, nous ne serions pas des êtres capables d'orienter éthiquement nos existences. L'éthique suppose la liberté, et paradoxalement, elle requiert, pour être,

la méta-éthique nous met en face de questions du genre : quel type de bien voulons-nous réaliser en promouvant l'essor de l'IA dans nos sociétés démocratiques ? Comment justifions-nous les valeurs que nous voulons retenir dans nos évaluations de l'IA ? Quelle vision du monde privilégier, quelle conception de l'homme et de ses fins, soutient-on en promouvant l'émergence d'une société numérique, composée d'intelligences artificielles ? Etc.

De telles questions fondamentales qui ressortent de la méta-éthique et d'un type de questionnement caractéristique de la philosophie morale, apparaissent peu présentes dans la structuration du champ institutionnel de l'éthique de l'IA et dans la façon de faire de l'éthique de l'IA. Un indice de cette indigence est la portion particulièrement congrue, voire inexistante, des financements publics consacrés à soutenir ce type de réflexions fondamentales, et pourtant essentielles pour une société démocratique et l'instauration d'une authentique démocratie technique. Il y a là en effet un enjeu d'auto-détermination politique : où et comment se décide, au nom de quelles raisons, de quelle représentativité démocratique et pour quelles fins, que l'avenir du travail, de la culture, voire plus largement de l'humanité doive nécessairement coïncider avec l'avenir de l'IA, ou avec tel ou tel type d'organisation technique à l'échelle sociétale ? Les normes, les conceptions du bien et du juste qui sous-tendent une telle organisation sont-elles justifiables?

Plutôt que de consacrer du temps et de l'argent à des travaux relevant de la philosophie morale fondamentale ou de la méta-éthique, les soutiens publics au développement de l'éthique de l'IA sont plutôt orientés en faveur du développement d'une éthique normative de l'IA, autour des enjeux que nous pose sa régulation et la nécessaire mise en place des processus de contrôle, des bonnes pratiques et des normes à instaurer pour garantir le développement d'une IA éthique, c'est-à-dire respectueuse de la dignité de la personne humaine, de ses droits et libertés fondamentales, mais aussi non-discriminatoire, non excluante et conforme aux principes et valeurs de nos régimes démocratiques. Ceci est heureux et bienvenu, mais qu'en est-il de la méta-éthique ? Cette dernière apparaît comme un parent pauvre de l'éthique de l'IA, dont les modalités relèvent surtout de l'éthique normative.

## *3.2. L'éthique normative*

A côté de la méta-éthique, l'éthique normative est le deuxième champ de la philosophie morale qu'il convient d'aborder. A la différence de la méta-éthique, l'éthique normative ne porte pas sur l'analyse des fondations et la justification ultime des concepts moraux. Elle soulève plutôt la question du *comment* de l'évaluation du caractère bon au mauvais, juste ou injuste des actions humaines. Elle recouvre ainsi un ensemble de théories qui se présentent comme des moyens pour juger du caractère bon ou mauvais, juste ou injuste des activités humaines. En s'appuyant *a priori* sur une certaine vision substantielle du bien et du juste (qui peut varier d'une théorie à l'autre), l'éthique normative propose des critères et un mode opératoire pour évaluer nos actions. Les théories classées sous la catégorie de l'éthique

---

la présence virtuelle de son non-être, un non-être de l'éthique sans cesse nié dans chacune de nos démarches et de nos actions morales.

normative se veulent donc plus efficaces, plus opératoires que la méta-éthique, quoi qu'elles restent encore très théoriques par rapport au champ des éthiques appliquées, troisième grand registre de la philosophie morale que nous abordons plus loin.

A la question du comment de l'évaluation du caractère bon au mauvais, juste ou injuste d'une action particulière, les éthiques normatives proposent de nombreuses réponses. Mais elles sont en général classées en 3 grandes catégories.

### 3.2.1. Les éthiques déontologistes ou principlistes

Une première forme de réponse relève des éthiques déontologistes ou principlistes (ces termes sont employés comme des synonymes en philosophie morale). Ces dernières soutiennent qu'une action est bonne ou juste si certaines règles ou principes qui fondent sa bonté ou sa justesse sont bien respectés. L'éthique des droits de l'homme, les codes de déontologie des professions, ou les éthiques religieuses, sont de bons exemples d'approches principlistes des questions éthiques.

### 3.2.2. Les éthiques conséquentialistes ou utilitaristes

À côté du déontologisme ou du principlisme en éthique, s'ajoute un second type de réponse que l'on appelle *conséquentialiste ou utilitariste*. Dans ce type de théorie domine l'idée que l'action la meilleure, éthiquement parlant, est celle qui produit les meilleures conséquences possibles, par exemple si suite à telle ou telle action ou politique privilégiée, le bonheur ou la santé du plus grand nombre sont maximisés.

Le déontologisme et le conséquentialisme présentent des formes de raisonnements moraux et de critères d'aide à la décision particulièrement sollicités en matière de santé publique, de recherche biomédical et d'éthique en matière d'informatique. La grande majorité des recommandations et des tentatives d'encadrement éthique de l'IA et du numérique parues ces dernières années relève ainsi d'approches déontologiques visant à énumérer les grands principes auxquels tout système d'IA devrait se soumettre pour garantir son « éthicité ».

Par ailleurs, les approches conséquentialistes, de même que les théories éthiques de type déontologistes, se prêtent toutes deux très bien à des projets de formalisation de l'éthique et de développements d'IA dites « éthiques » : les premières parce qu'elles correspondent dans leur mode opératoire à des formes de calcul en termes de balance bénéfices/risques; les secondes parce qu'elles conduisent à l'établissement de listes de principes desquels peuvent s'inférer un ensemble d'implications logiques dans le champs pratique. Le déontologisme et le conséquentialisme font ainsi la part belle à la pensée déductive. Les raisonnements qu'ils privilégient entretiennent une relation d'isomorphie avec les modes opératoires d'une pensée axiomatique, calculatoire, et donc algorithmisable. Ce sont donc ces éthiques-là que l'on présuppose quand on parle en général d'apprentissage de l'éthique par des IA ou de projet de produire des IA ou des robots doués de capacités de raisonnement moral.

La voiture autonome est un exemple bien connu d'innovation en matière d'IA pour lequel on invoque la nécessité d'un traitement éthique automatisé des données de route, en particulier quand des vies sont exposées. Mais avec quel type de théorie morale entraîner ou programmer une machine intelligente ? Il n'est pas surprenant de constater qu'un tour de la littérature[11] montre rapidement que les deux théories, systématiquement envisagées et débattues pour programmer une IA éthique dans des voitures autonomes, relèvent soit du conséquentialisme – avec sa version dominante, l'utilitarisme –, soit du déontologisme. Dans le premier cas, la programmation d'une voiture autonome devrait suivre la règle de la maximisation du bien-être du plus grand nombre, et, au terme d'un calcul d'optimisation fondé sur une valeur d'utilité numérique attribuée à chaque résultat possible, une voiture utilitariste privilégiera systématiquement, dans une situation d'accident inévitable, le scénario dont le résultat du calcul des dommages matériels et humains sera le plus faible. Dans le second cas, celui du raisonnement déontologique, les comportements préprogrammés d'une voiture autonome en cas d'accident seront moralement justifiés selon leur conformité ou non à certains principes absolus, c'est-à-dire indépendants des circonstances, comme l'ordre par exemple de préserver en toute situation la vie des occupants de la voiture, ou encore la règle, au nom de l'égalité et de la dignité inconditionnelle de toute personne humaine, de ne jamais désigner au terme d'un calcul des personnes qui doivent être épargnées et d'autres qui doivent être sacrifiées. C'est la position, par exemple, d'un chercheur comme Alexei Grinbaum[12] qui soutient qu'en cas de dilemme moral, une voiture autonome devrait laisser au hasard le fin mot de l'histoire, car l'idée que nos destinées soient prédéterminées, en amont de ce qu'il pourrait nous arriver, par les opérations d'une machine, est insupportable.

Au-delà du caractère insatisfaisant de ces deux solutions utilitaristes ou déontologistes, dans lesquelles nous enferme le fameux dilemme du tramway appliqué au cas de la voiture autonome, c'est cet enfermement même dans un tel dilemme qui soulève un problème plus général : la voiture autonome ne suscite-t-elle pas d'autres questions, d'ordre méta-éthique par exemple, que celles que nous pose une situation aporétique d'accident inévitable ? La réduction de la réflexion éthique au débat entre IA utilitariste ou IA déontologiste risque encore d'oblitérer un second problème plus général. En effet, toute entreprise de programmation d'une IA éthique ne peut éviter un élément d'arbitrage et de préférence humaine à son principe. Il s'agit par exemple de la préférence des producteurs d'une voiture autonome en faveur de l'utilitarisme ou du déontologisme, car ces éthiques normatives se prêtent plus naturellement à leur formalisation en langage algorithmique que d'autres types de théories morales. Ce sont encore des préférences et des décisions, prises par certains acteurs, qui concourent à déterminer et pré-programmer les critères d'attributions de valeurs (sous forme de points) qu'appliquera, en cas d'accident par exemple, une machine utilitariste pour pondérer (évaluer) les éléments d'une situation et discriminer entre les vies qui doivent être sauvées et celles qui seront sacrifiées. Dans le cas d'une voiture équipée d'une intelligence artificielle déontologiste, ce sont toujours des préférences humaines particulières qui président, en amont des situations critiques, au choix des principes et valeurs qui doivent déterminer le comportement du véhicule, indépendamment de l'appréciation morale des acteurs impliqués dans ces situations lorsqu'elles surviennent..

---

[11] Voir V. Shaffner, "Caught Up in Ethical Dilemmas: An Adapted Consequentialist Perspective on Self-Driving Vehicles", *Frontiers in Artificial Intelligence and Applications*, Vol. 311, 2018, pp. 327-335.

[12] A. Grinbaum, *Les robots et le mal*, Paris, Desclée de Brouwer, 2019.

Toute préférence ou choix traduit sous forme d'algorithme relève de la décision et de la responsabilité humaines, et appelle par conséquent un débat sur sa justification. Certaines personnes morales ou groupes d'individus peuvent-ils incarner leur vision particulière du bien et du juste dans des infrastructures technologiques susceptibles d'impacter l'ensemble d'une société, sans qu'un débat démocratique ne s'impose préalablement quant aux conditions et aux conséquences de telles opérations ? Pourquoi tel système d'attribution de valeur et non tel autre ? Pourquoi tel principe serait-il considéré comme absolu, quand d'autres ne seraient que relatifs ? Nous retrouvons ici un besoin de réflexion éthique sur l'éthique de l'IA des voitures autonomes, autrement dit un besoin de recul et de méta-éthique. Cette réflexion existe dans la littérature académique, mais elle paraît peu organisée au niveau des institutions publiques et du débat démocratique. Nous avons pourtant besoin d'une démocratie technique. C'est un nouveau besoin dans l'histoire de nos sociétés démocratiques, mais il me semble essentiel pour cultiver l'esprit d'une démocratie.

### 3.2.3. Les éthiques des vertus

Après avoir abordé les approches déontologiques et conséquentialistes, la troisième grande réponse des éthiques normatives au comment de l'évaluation du caractère bon ou mauvais, juste ou injuste des actions humaines, relève de *l'éthique des vertus*. Les penseurs des éthiques de la vertu considèrent qu'une action est morale quand elle repose sur des vertus et permet de les développer. Par vertu, on peut entendre des dispositions du caractère socialement fécondes, telles que la justice, la tempérance et le courage, la magnanimité, la prudence ou la bonté, etc., par oppositions aux vices, tels que l'avidité, la lâcheté, l'avarice, l'intolérance, etc. Dans ce troisième grand champ théorique de l'éthique normative, une action sera considérée comme bonne ou juste, si son agent est vertueux. Le critère de bonté ou d'équité d'une action ne porte pas ici sur une adéquation à des principes (déontologisme) ou sur ses conséquences bonnes ou mauvaises, justes ou injustes (utilitarisme), mais sur la qualité, la vertu, reconnue, ou pas, de l'agent. Nous sommes ici dans les éthiques de l'exemple et de l'exemplarité, de la personne providentielle, et des éthiques fondées sur la réputation.

Associée au champ de l'IA, l'éthique des vertus constitue une ressource de la tradition morale occidentale peu sollicitée dans la littérature et nos débats publics en matière d'IA. Pourtant, ce champ de la pensée morale est particulièrement vaste. Le champ de la robotique sociale y fait toutefois référence, dès lors qu'un des enjeux de ce domaine d'étude est de développer des robots qui puissent apprendre et simuler des comportements humains – de préférence des comportements vertueux, socialement féconds et appréciables. Il reste toutefois que l'apprentissage artificiel des vertus s'expose à un obstacle redoutable : une vertu est une disposition à agir de façon excellente en fonction de circonstances sociales toujours singulières, qui demandent de la part de l'agent une faculté d'appréciation fine des enjeux multiples d'une situation. Il ne s'agit pas d'une capacité indépendante d'un contexte d'action. Par exemple, on ne peut dire que le courage est « en soi » une vertu, car il était certainement plus vertueux pour un nazi d'être un déserteur et un lâche, que d'agir avec détermination et courage pour son Führer. La qualité morale d'une disposition acquise au cours de la vie, son caractère vertueux ou pas, dépendent des intentions et des circonstances de sa mise en œuvre. La qualification d'une disposition pratique comme vertu ou vice dépend de ses conséquences et de sa signification dans un contexte d'action qui s'avère toujours particulier. Un autre champ de l'IA, où l'éthique des vertus pourrait être explorée, pourrait se situer au niveau des acteurs de l'IA. Doit-on attendre des acteurs de l'IA qu'ils développent certaines vertus éthiques (mais aussi épistémiques, c'est-à-dire relatives à la connaissance) dont on pourrait, à l'analyse, souligner l'importance dans leurs pratiques et leurs responsabilités ?

## *3.3. L'éthique appliquée*

Le troisième champ de l'éthique, après celui de la méta-éthique et des éthiques normatives, est celui de l'éthique appliquée (appelée aussi éthique pratique). Ce champ, tourné clairement vers les terrains concrets de l'éthique et les questions sociétales, est partagé par deux grandes tendances. Un bon nombre d'*éthiques appliquées* privilégient une approche *top-down* des terrains en se donnant pour projet d'appliquer telle ou telle théorie de l'éthique normative à des cas pratiques. D'autres formes d'éthiques pratiques valorisent plutôt les approches *down-top*, comme les *éthiques pragmatistes* ou les *éthiques contextuelles*. Dans ces approches, la résolution des problèmes éthiques doit tenir compte des spécificités des contraintes factuelles et normatives de contexte toujours spécifiques. La démarche éthique implique alors un travail d'enquête, d'explicitation et d'élucidation des enjeux normatifs d'une situation, qui bien souvent suppose l'implication des acteurs directement concernés dans la formulation du problème éthique et la mise en œuvre des étapes concrètes de sa résolution.

## 4. Pour une éthique humaine de l'IA, pragmatiste et située

Par rapport aux approches classiques des questions éthiques qui relèvent des éthiques normatives, ou par rapport au champ des éthiques pratiques dont la règle est d'appliquer la théorie et les raisonnements d'une forme ou l'autre d'éthique normative à des cas ou des situations particulières, les éthiques pragmatistes et contextuelles entretiennent un rapport instrumental avec les traditions morales et les grandes théories éthiques occidentales. Elles partent du principe que la résolution d'un problème éthique implique de prendre en compte un ensemble particulièrement riche d'informations et de registres de normativités sociale, juridique, morale, culturelle, etc., qui peuvent apparaître conflictuels dans une situation éthiquement problématique. Résoudre une telle situation implique un travail de fond, qui ne peut se réduire à la simple application d'une théorie morale particulière, et qui impliquera le plus souvent des choix, des interprétations, et la création de nouvelles règles ou normes dans le but de résoudre le problème levé.

L'important pour les éthiques pragmatistes et contextuelles n'est donc pas la cohérence d'un système ou d'une théorie à mettre en œuvre de façon procédurale, mais de résoudre des problèmes concrets rencontrés dans les institutions. Ces problèmes relèvent le plus souvent du dilemme, du paradoxe, ou de conflits entre des registres de normativités qui ne s'accordent pas toujours. On les qualifie de « problème noueux » (*wicked problems*[13]) car ils demandent en général de trouver des solutions non déductibles de raisonnements linéaires, de principes ou de formules de résolution *a priori*. Dans ces cas de figure, l'intervenant en éthique œuvre avec les parties prenantes qui l'interpellent, à la mise à place des meilleures conditions organisationnelles, sociales et techniques au sein desquels des publics concernés vont être positionnés en situation d' « empouvoirement » (*empowerment*) éthique (comme le traduisent nos collègues québécois) au sens où leurs capacités de réflexion, d'analyse, de méthode, et de délibération collective vont pouvoir se développer et produire collectivement des solutions innovantes à des problèmes rencontrés.

---
[13] H. Rittel and M. Webber, "Dilemmas in a General Theory of Planning", *Policy Sciences*, Vol. 4, 1973, pp. 155-169.

Si l'éthique pragmatique et contextuelle s'est développée depuis de nombreuses années dans les champs de la santé et du social, elle apparaît encore bien trop peu mobilisée dans le champ de l'IA. Or, c'est une pratique collective de l'éthique, de façon pragmatique et contextuelle, dans les entreprises ou dans les institutions publiques, qui peut permettre de faire l'expérience de ce qui fait véritablement la spécificité de la dimension humaine de l'éthique, comme pouvoir d'interprétation, d'évaluation et d'invention normative. Nous retrouvons là le sens et le contenu premiers de l'éthique dont nous avons proposé l'explicitation dans les premières sections de notre étude.

Nous disposons, en tant qu'être humain, d'un pouvoir dynamique de création de normes qui s'expriment et se traduisent au plan moral, juridique, déontologique, social, etc. Mais ce pouvoir s'accompagne aussi d'une capacité d'interprétation, de jugement, de réforme et de subversion normative qui doit être bien distingué de tous les systèmes particuliers de règles et de principes, lesquels ne sont jamais que les conséquences de ce pouvoir. Ce pouvoir nous distingue des machines et nous responsabilise au plus haut point quant aux sens et aux finalités de nos actions, dès lors qu'il dépend de nous, de façon radicale, d'instaurer et de nous porter garants des conditions normatives d'une vie bonne, avec et pour autrui, dans des institutions justes – pour reprendre ici la définition de la petite éthique de Paul Ricoeur. Il serait bien sûr désastreux de renoncer à ce pouvoir et cette responsabilité, en nous laissant prendre par l'illusion de croire que des machines pourraient être éthiques.